\documentclass[9pt,twocolumn,twoside]{opticajnl}

\journal{ol} 

\setboolean{shortarticle}{true}

\usepackage{multirow}

\title{A Silicon Nitride Microring Modulator for High-Performance Photonic Integrated Circuits}

\author{Venkata Sai Praneeth Karempudi}
\author{Ishan G Thakkar}
\author{Jeffrey Todd Hastings}

\affil[1]{Department of Electrical and Computer Engineering, University of Kentucky, Lexington, Kentucky, USA, 40508}

\affil[*]{Corresponding author: kvspraneeth@uky.edu}

\begin{abstract}
The use of the Silicon-on-Insulator (SOI) platform has been prominent for realizing CMOS-compatible, high-performance photonic integrated circuits (PICs). But in recent years, the silicon-nitride-on-silicon-dioxide (SiN-on-SiO$_2$) platform has garnered increasing interest as an alternative, because of its several beneficial properties over the SOI platform, such as low optical losses, high thermo-optic stability, broader wavelength transparency range, and high tolerance to fabrication-process variations. However, SiN-on-SiO$_2$ based active devices, such as modulators, are scarce and lack in desired performance due to the absence of free-carrier-based activity in the SiN material and the complexity of integrating other active materials with SiN-on-SiO$_2$ platform. This shortcoming hinders the SiN-on-SiO$_2$ platform for realizing active PICs. To address this shortcoming, in this article, we demonstrate a SiN-on-SiO$_2$ microring resonator (MRR) based active modulator. Our designed MRR modulator employs an Indium-Tin-Oxide (ITO)-SiO$_2$-ITO thin-film stack as the active upper cladding and leverages the free-carrier assisted, high-amplitude refractive index change in the ITO films to affect a large electro-refractive optical modulation in the device. Based on the electrostatic, transient, and finite difference time domain (FDTD) simulations, conducted using photonics foundry-validated tools, we show that our modulator achieves 450 pm/V resonance modulation efficiency, $\sim$46.2 GHz 3-dB modulation bandwidth, 18 nm free-spectral range (FSR), 0.24 dB insertion loss, and 8.2 dB extinction ratio for optical on-off-keying (OOK) modulation at 30 Gb/s. 
\end{abstract}

\begin{document}

\maketitle

\section{Introduction}
Driven by the rise of CMOS-compatible processes for fabricating photonic devices, photonic integrated circuits (PICs) are inexorably moving from the domain of long-distance communications to chip-chip and even on-chip applications. It is common for the PICs to incorporate optical modulators to enable efficient manipulation of optical signals, which is a necessity for the operation of active PICs. Recent advances in the CMOS-compatible silicon-on-insulator (SOI) photonic platform has fundamentally improved the applicability of SOI PICs \cite{chrostowski2014}, \cite{bogaerts2020}, \cite{harris2018}. But in the last few years, the silicon-nitride-on-silicon-dioxide (SiN-on-SiO$_2$) platform has gained tremendous attention for realizing PICs. This is because the SiN-on-SiO$_2$ platform has several advantages over the SOI platform. Compared to silicon (Si), the SiN material has a much broader wavelength transparency range (500nm-3700nm), smaller thermo-optic coefficient, and lower refractive index \cite{wilmart2019}. The lower refractive index of SiN means that SiN offers smaller index contrast with SiO$_2$ compared to Si. This in turn makes the SiN-on-SiO$_2$ based monomode passive devices (e.g., waveguides, microring resonators (MRRs)) less susceptible to \textit{(i)} propagation losses due to the decreased sensitivity to edge roughness \cite{bauters2011ultra}, and \textit{(ii)} aberrations in the realized device dimensions due to fabrication-process variations \cite{wilmart2019}. In addition, the smaller thermo-optic coefficient of SiN makes it possible to design nearly athermal photonic devices using SiN \cite{gao2017silicon}. Moreover, SiN devices and circuits exhibit increased efficiency of nonlinear parametric processes compared to Si \cite{levy2011integrated}.

Despite these favorable properties of the SiN-on-SiO$_2$ platform, SiN-on-SiO$_2$ based active devices such as modulators are scarce and lack in free spectral range (FSR), modulation bandwidth, and modulation efficiency  \cite{alexander2018nanophotonic}. The lack in efficiency is because of the lack of the free-carriers based activity in the SiN material and the general difficulty of incorporating other active materials with the SiN-on-SiO$_2$ platform. This in turn limits the use of the SiN-on-SiO$_2$ platform to realizing only passive PICs. To overcome this shortcoming, there is impetus to heterogeneously integrate active photonic materials and devices with SiN-on-SiO$_2$ passive devices (e.g., \cite{goyvaerts2021, phare2015graphene,jin2018piezoelectrically,ahmed2019high,hermans2019integrated,wang2022,ahmed2019,liu2020,wang2022ultra}). When such efforts of integrating electro-optically active materials with the SiN-on-SiO$_2$ platform come to fruition, it will be possible to design high-performance and energy-efficient SiN-on-SiO$_2$ based active and passive PICs. 

Different from such prior efforts, in this article, we demonstrate the use of the high-amplitude electro-refractive activity of Indium-Tin-Oxide (ITO) thin films to realize a SiN-on-SiO$_2$ based optical on-off-keying (OOK) modulator. We show, based on the electrostatic, transient, and finite difference time domain (FDTD) simulations conducted using the photonics foundry-validated tools from Lumerical/Ansys, that our modulator achieves 450 pm/V resonance modulation efficiency, $\sim$46.2 GHz 3-dB modulation bandwidth, 18 nm free-spectral range (FSR), 0.24 dB insertion loss, and 8.2 dB extinction ratio for optical OOK modulation at 30 Gb/s. \textit{Based on the obtained simulation results, we advocate that our modulator can achieve better performance compared to the existing SiN modulators from prior works}.

\section{Related Work and Motivation}
A plethora of Si and SiN based integrated optical modulator designs have been formulated in prior works \cite{rahim2021}. But among these modulator designs, MRR based modulators have gained widespread attention due to their high wavelength selectivity, compact size, and compatibility for cascaded dense wavelength division multiplexing (DWDM).
Recently several MRR based SiN-on-SiO$_2$ modulators have also been demonstrated (e.g., \cite{phare2015graphene,alexander2018nanophotonic,jin2018piezoelectrically,ahmed2019high,hermans2019integrated,wang2022,ahmed2019,liu2020,wang2022ultra}). In \cite{phare2015graphene}, a graphene integrated electro-optic SiN MRR modulator has been reported. In \cite{ahmed2019high}, a hybrid SiN-LiNbO$_3$ platform based racetrack resonator modulator has been presented. Similarly, SiN modulators based on lead zinconate titanate and zinc oxide/zinc sulphide as active materials are demonstrated in \cite{alexander2018nanophotonic} and \cite{hermans2019integrated}. In \cite{jin2018piezoelectrically}, a SiN modulator that achieves tuning via photo-elastic effect has been demonstrated. Compared to these modulator designs from prior works, we present a different, ITO-based electro-refractive SiN-on-SiO$_2$ modulator that achieves substantially better modulation bandwidth, efficiency, and FSR.  

\begin{figure}[h!]
    \centering
    \includegraphics[scale = 0.1]{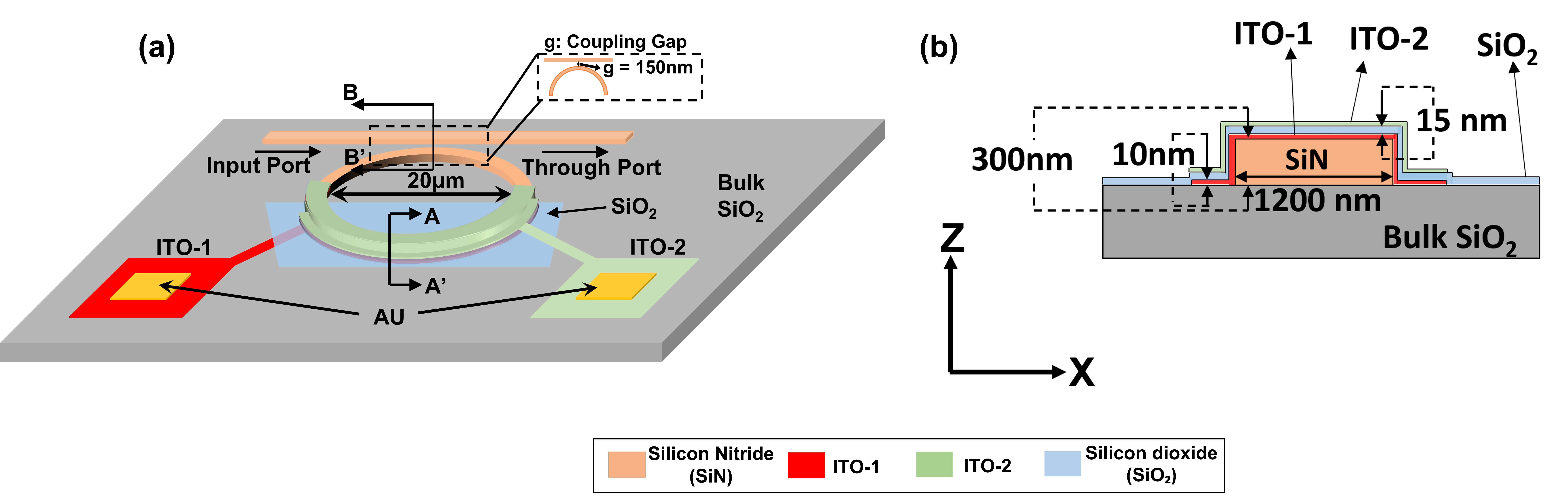}
    \caption{(a) Top view, (b) Cross-sectional view (along AA') of our SiN-on-SiO$_2$ MRR modulator.}
    \label{Fig:1}
\end{figure}

\section{Design of our SiN-on-SiO$_2$ Modulator}
In this section, firstly we describe the structure and operating principle of our modulator design. Then, we discuss the characterization results for our modulator that we have obtained through photonics foundry-validated simulations. We also compare our modulator with several SiN based MRR modulators from prior works, in terms of modulation bandwidth, modulation efficiency, and FSR.

\begin{figure} [h!]
    \includegraphics[scale = 0.26]{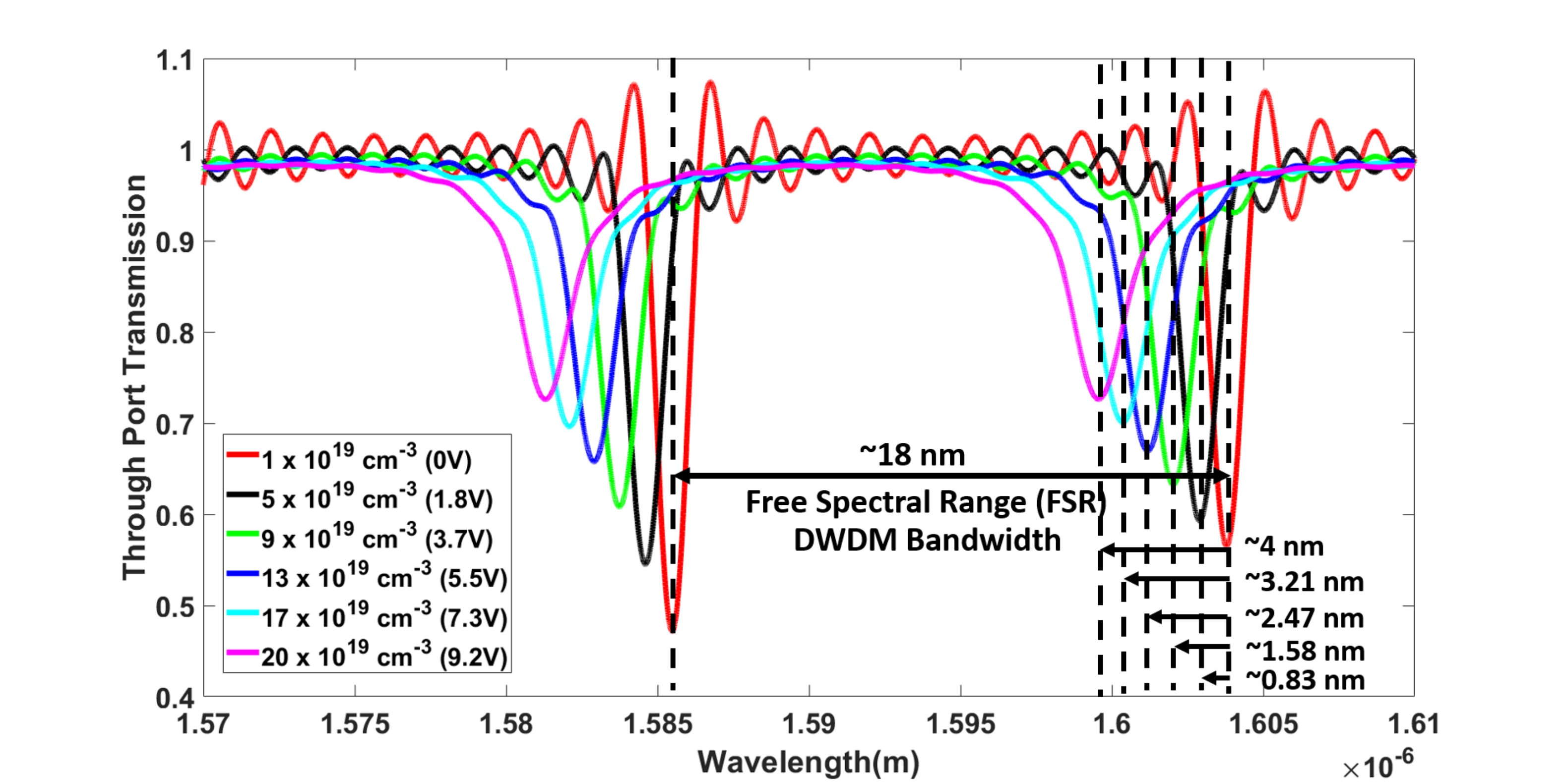}
    \caption{Transmission spectra of our modulator.}
    \label{Fig:2}
\end{figure}


\subsection{Structure and Operating Principle}
Fig. \ref{Fig:1}(a) and Fig. \ref{Fig:1}(b), respectively, show the top-view and cross-sectional schematic of our SiN-on-SiO$_2$ MRR modulator. The active region in the upper cladding of the modulator consists of a stack of two indium tin oxide (ITO) thin films with a silicon dioxide (SiO$_2$) thin film in between (an ITO-SiO$_2$-ITO stack). From Fig. \ref{Fig:1}(b), we have a 300 nm thick SiN-based MRR waveguide, two 10 nm thick ITO films, and 15 nm thick SiO$_2$ layer. Upon applying voltage across the ITO-SiO$_2$-ITO stack (through the Au pads shown in Fig. \ref{Fig:1}(a)), free carriers accumulate in the ITO films at the ITO-SiO$_2$ interfaces for up to ~5 nm depth in the ITO films \cite{chrostowski2014}, making these accumulation regions in the ITO films high-carrier-density active regions. In these regions, a free-carriers-assisted, large-amplitude modulation in the permittivity and refractive index of the ITO material has been previously reported \cite{chrostowski2014}. We evaluate this free-carriers based index modulation in the ITO films using the Drude-Lorentz model from \cite{ma2015indium}. Accordingly, as the carrier concentration in the ITO accumulation regions increases, the refractive index of the ITO films decreases. As a result, the effective refractive index of our SiN-on-SiO$_2$ modulator design from Fig. \ref{Fig:1} also decreases, causing a blue shift in its resonance wavelength that in turn causes a transmission modulation at the through port of the modulator. The electro-refractive activity of our SiN-on-SiO$_2$ MRR modulator design is confined only in the ITO-SiO$_2$-ITO cladding. This is different from the Si-SiO$_2$-ITO capacitor based MRR modulator from \cite{li2019silicon}, which has the electro-refractive as well as electro-absorptive activities in both its Si-based MRR core and SiO$_2$-ITO based cladding.


\begin{table}[]
\caption{Free-carrier concentration (N), real index (Re($\eta_{ITO}$)), and imaginary index (Im($\eta_{ITO}$)) for the ITO accumulation layer in our modulator. The real and imaginary effective index (Re($\eta_{eff}$), Im($\eta_{eff}$)), operating voltage (V), and induced resonance shift ($\Delta\lambda_{r}$) for our modulator.}
\begin{tabular}{|c|c|c|c|c|c|c|}
\hline
\begin{tabular}[c]{@{}c@{}}N \\ ($cm^{-3}$)\end{tabular} & \begin{tabular}[c]{@{}c@{}}Re\\ ($\eta_{ITO}$)\end{tabular} & \begin{tabular}[c]{@{}c@{}}Im\\ ($\eta_{ITO}$)\end{tabular} & \begin{tabular}[c]{@{}c@{}}Re\\ ($\eta_{eff}$)\end{tabular} & \begin{tabular}[c]{@{}c@{}}Im\\ ($\eta_{eff}$)\end{tabular} & V   & \begin{tabular}[c]{@{}c@{}} $\Delta\lambda_{r}$\\ (pm)\end{tabular} \\ \hline
1 × $10^{19}$                                            & 1.9556                                              & 0.0100                                              & 1.9735                                              & 0.0001                                              & 0   & 0                                                  \\ \hline
5 × $10^{19}$                                            & 1.9111                                              & 0.0403                                              & 1.9724                                              & 0.0003                                              & 1.8 & 830                                                \\ \hline
9 × $10^{19}$                                             & 1.8667                                              & 0.0896                                              & 1.9712                                              & 0.0006                                              & 3.7 & 1580                                               \\ \hline
13 × $10^{19}$                                           & 1.8222                                              & 0.1289                                              & 1.9701                                              & 0.0011                                              & 5.5 & 2470                                               \\ \hline
17 × $10^{19}$                                            & 1.7778                                              & 0.1582                                              & 1.9692                                              & 0.0017                                              & 7.3 & 3210                                               \\ \hline
20 × $10^{19}$                                            & 1.7333                                              & 0.1874                                              & 1.9680                                              & 0.0022                                              & 9.2 & 4000                                               \\ \hline
\end{tabular}
\label{Table:1}
\end{table}

\subsection{Simulations Based Characterization}
We performed electrostatic simulations of our ITO-SiO$_2$-ITO thin-film stack based SiN-on-SiO$_2$ modulator in the CHARGE tool of DEVICE suite from Lumerical \cite{lumericalwebsite}, to evaluate the required voltage levels across the Au pads (Fig. \ref{Fig:1}(a)) for achieving various free-carrier concentrations in the ITO films. Then, based on the Drude-Lorentz dispersion model from \cite{ma2015indium}, we extracted the corresponding ITO index change values for various free-carrier concentrations. These results are listed in Table \ref{Table:1}. Using these index values from Table \ref{Table:1}, we modeled our MRR modulator in the MODE tool from Lumerical \cite{lumericalwebsite} for finite-difference-time-domain (FDTD) and finite-difference eigenmode (FDE) analysis. For this analysis, we used the Kischkat model \cite{kischkat2012} of stoichiometric silicon nitride to model the MRR device. From this analysis, we extracted the effective index change and transmission spectra of our modulator (shown in Table \ref{Table:1} and Fig. \ref{Fig:2} respectively) at various applied voltages for the operation around 1.6 $\mu$m wavelength (L-band). From Fig. \ref{Fig:2}, our modulator achieves up to 4 nm resonance shift upon applying 9.2 V across the thin-film stack, which renders the resonance tuning (modulation) efficiency of $\sim$450 pm/V. This is a crucial outcome as our MRR modulator has relatively very low overlap between the optical mode and free-carrier perturbation (only ~10\% of the guided optical mode overlaps with the ITO-based upper cladding) compared to the silicon ITO-based modulators (e.g., \cite{li2019silicon}).

\begin{figure}[h!]
    \centering
    \includegraphics[scale = 0.075]{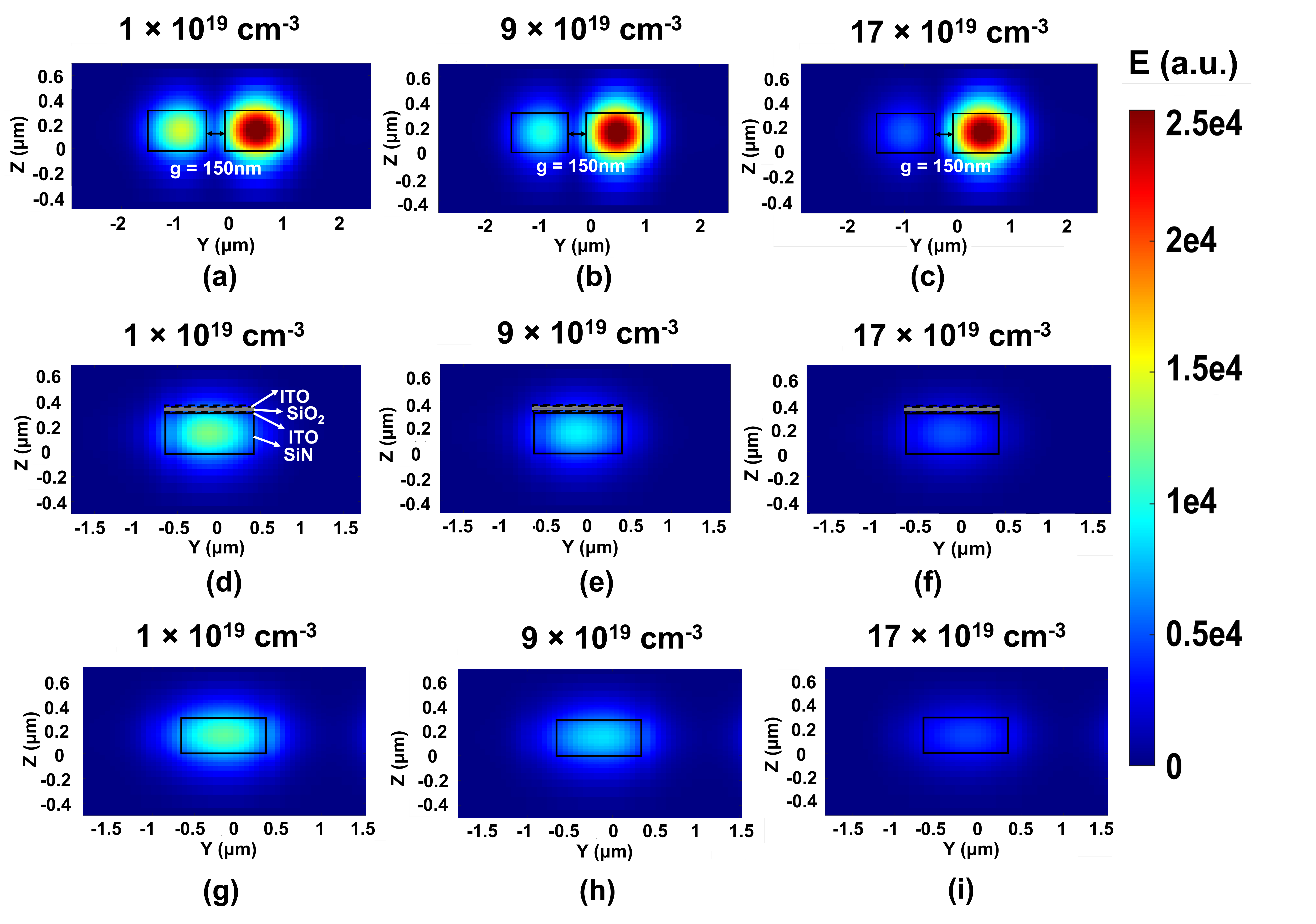}
    \caption{Cross-sectional electric-field profiles of the fundamental TE mode evaluated at the coupling section (along BB'in Fig. \ref{Fig:1}(a)) ((a)-(c)), across the rim (along AA' in Fig. \ref{Fig:1}) ((d)-(f)), and at the through port of our SiN-on-SiO$_{2}$ MRR modulator ((g)-(i)), for three different free-carrier concentrations of ITO (Table \ref{Table:1}) namely 1$\times$10$^{19}$ cm$^{-3}$ (for (a),(d),(g)), 9$\times$10$^{19}$ cm$^{-3}$  (for (b),(e),(f)), and  17$\times$10$^{19}$ cm$^{-3}$  (for (c),(f),(i)), using the variational FDTD (varFDTD) solver \cite{lumericalwebsite}.}
    \label{Fig:3}
\end{figure}

Fig. \ref{Fig:3} illustrates the cross-sectional electric-field profiles of the fundamental TE mode evaluated for three different free-carrier concentrations of ITO, namely 1$\times$10$^{19}$ cm$^{-3}$, 9$\times$10$^{19}$ cm$^{-3}$, and 17$\times$10$^{19}$ cm$^{-3}$, at three different cross-sectional regions of our MRR modulator. To evaluate these profiles, we used the variational FDTD (varFDTD) solver of the MODE tool of the DEVICE suite from ANSYS/Lumerical. In fact, Fig. \ref{Fig:3} shows a 3$\times$3 grid of field profiles. Each row in this grid corresponds to field profiles collected for a particular cross-sectional region of our MRR modulator across different free-carrier concentrations of ITO.  Similarly, each column in the grid corresponds to field profiles collected for a particular free-carrier concentration of ITO across three different regions of the modulator.

As per the discussion in the previous subsection, the increase in the free-carrier concentration in the ITO layers caused due to the increase in the applied bias across the ITO-SiO$_{2}$-ITO stack, decreases the effective index of our modulator. This in turn induces a blue shift in the resonance wavelength of our modulator. Due to this blue shift in the resonance wavelength, the amount of optical power coupled from the input port into the MRR cavity at the coupling region decreases. This can be clearly observed from the field profiles collected at the coupling region along BB’; as the free-carrier concentration increases from Fig. \ref{Fig:3}(a) to Fig. \ref{Fig:3}(c), the intensity of the coupled field in the MRR at the cross-section BB' also 
decreases. The decrease in the coupled field intensity at BB' naturally results in the decrease of the steady-state field intensity inside the MRR waveguide. As a result, at the cross-section AA', the field intensity can be observed to decrease with the increase in the free-carrier concentration in the ITO layers, as we move from Fig. \ref{Fig:3}(d) to Fig. \ref{Fig:3}(f) in the middle row of Fig. \ref{Fig:3}. Atop the steady-state field intensity inside the MRR cavity, the field intensity at the through port (hence, the output optical power at the through port) of the MRR also decreases naturally with the increase in the free-carrier concentration. This can be observed in the bottom row of Fig. \ref{Fig:3}. The modulation of the optical output power at the through port with the change in the free-carrier concentration in the ITO layers corroborates the electro-refractive activity of our modulator. 

In addition, as we move from the top row (coupling region field profiles) to the bottom row (through port field profile) within each column in Fig. \ref{Fig:3}, the field intensity slightly decreases. This provides evidence that, for each column (i.e., for each specific free-carrier concentration), the optical field intensity undergoes optical-loss-induced attenuation as the light waves travel along the propagation path from the coupling region (top row) to the through port (bottom row).

\begin{figure}[h!]
    \centering
    \includegraphics[scale = 0.1]{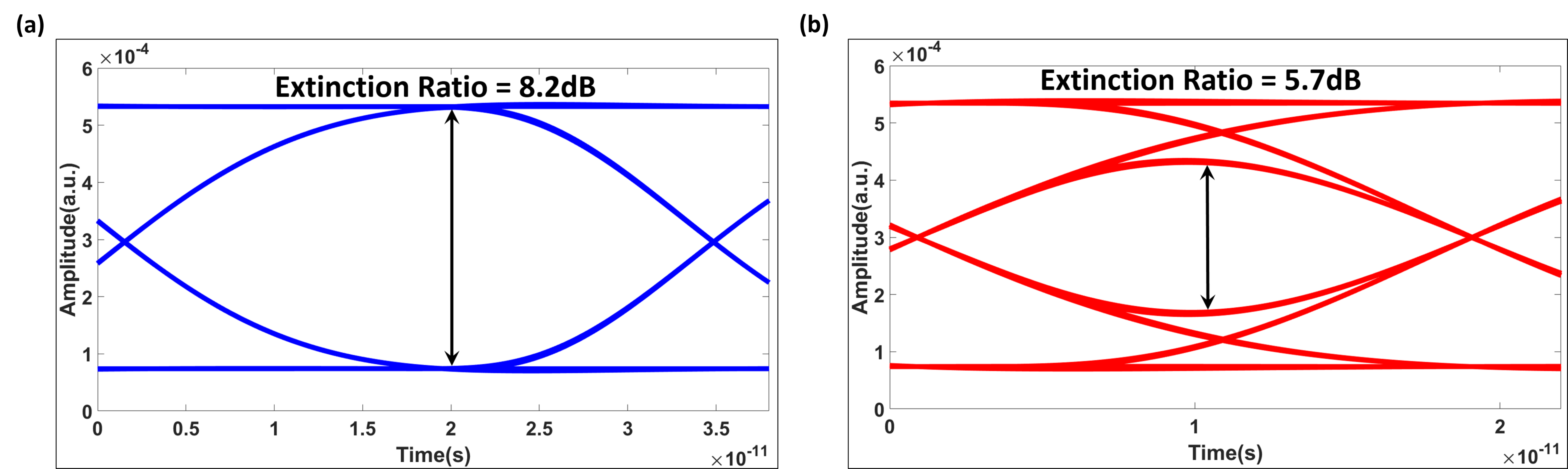}
    \caption{Optical eye diagrams for (a) 30 Gb/s and (b) 55 Gb/s OOK inputs to our modulator.}
    \label{Fig:4}
\end{figure}

Further, from the spectra in Fig. \ref{Fig:2}, we evaluate the FSR of our modulator to be $\sim$18 nm. We evaluated (using the Lumerical MODE tool) the insertion loss and loaded Q-factor of our modulator to be $\sim$0.235 dB and $\sim$2000 respectively. We also evaluated the capacitance density of the ITO thin-film stack covering the MRR rim (using the Lumerical CHARGE tool) to be $\sim$2.3 fF$/\mu$m$^{2}$ for the 15 nm thick SiO$_2$ layer. Moreover, we modeled our modulator in Lumerical INTERCONNECT, to simulate optical eye diagrams for the modulator at 30 Gb/s and 55 Gb/s operating bitrates (Fig. \ref{Fig:4}). As evident (Fig. \ref{Fig:4}), our modulator can achieve 8.2 dB extinction ratio for OOK modulation at 30 Gb/s bitrate.

\begin{table}[h!]
\caption{Modulation bandwidth (MB) (optical (O) and Electrical (E)), modulation efficiency (ME), FSR and energy efficiency (EE) corresponding to various SiN based MRR modulators (modulator type (MT)) from prior works obtained from simulations (*) and experiments, compared with our simulated SiN-on-SiO$_2$ MRR modulator.}
\begin{tabular}{|c|cc|c|c|c|}
\hline
\multirow{2}{*}{\begin{tabular}[c]{@{}c@{}}MT
\end{tabular}} & \multicolumn{2}{c|}{MB}                                                                                                       & \multirow{2}{*}{\begin{tabular}[c]{@{}c@{}}ME   \\    (pm/V)\end{tabular}} & \multirow{2}{*}{\begin{tabular}[c]{@{}c@{}}FSR\\   (nm)\end{tabular}} & \multirow{2}{*}{\begin{tabular}[c]{@{}c@{}}EE\\  (pJ/bit)\end{tabular}} \\ \cline{2-3}
                                                                           & \multicolumn{1}{c|}{\begin{tabular}[c]{@{}c@{}}O-MB \\   (GHz)\end{tabular}} & \begin{tabular}[c]{@{}c@{}}E-MB\\ (GHz)\end{tabular} &                                                                            &                                                                       &                                                                         \\ \hline
\cite{wang2022}                                           & \multicolumn{1}{c|}{0.06}                                                 & 0.02                                              & 1.6                                                                        & 4×10$^{-3}$                                                                & 1×10$^{-3}$                                                                  \\ \hline
\cite{ahmed2019}                                          & \multicolumn{1}{c|}{1.7}                                                  & N/A                                               & 0.01*                                                                      & 0.58                                                                  & N/A                                                                     \\ \hline
\cite{liu2020}                                            & \multicolumn{1}{c|}{0.01}                                                 & 7.9                                               & 1                                                                          & 113                                                                   & 3.7×10$^{4}$                                                                 \\ \hline
\cite{wang2022ultra}                                      & \multicolumn{1}{c|}{0.03}                                                 & 0.03                                              & 1.6                                                                        & 0.3                                                                   & 1×10$^{-3}$                                                                  \\ \hline
\cite{phare2015graphene}                                  & \multicolumn{1}{c|}{161}                                                  & 30                                                & 100*                                                                       & 4.7                                                                   & 0.8                                                                     \\ \hline
\cite{alexander2018nanophotonic}                          & \multicolumn{1}{c|}{87}                                                   & 35.6                                              & 67*                                                                        & 1.74                                                                  & N/A                                                                     \\ \hline
\cite{jin2018piezoelectrically}                           & \multicolumn{1}{c|}{0.19}                                                 & 1.3×10$^{-3}$                                          & 137.5*                                                                     & 2.2*                                                                  & 0.11                                                                    \\ \hline
\cite{ahmed2019high}                                      & \multicolumn{1}{c|}{25}                                                   & N/A                                               & 2.9                                                                        & 0.21*                                                                 & N/A                                                                     \\ \hline
\cite{hermans2019integrated}                              & \multicolumn{1}{c|}{1.55}                                                 & 5.9                                               & 0.2                                                                        & 0.6                                                                   & 53                                                                      \\ \hline

\cite{zhao2022}                              & \multicolumn{1}{c|}{1.77}                                                 & N/A                                               & 5.8                                                                        & 0.4*                                                                   & N/A                                                                      \\ \hline

\begin{tabular}[c]{@{}c@{}}Ours \end{tabular}                    & \multicolumn{1}{c|}{93.62*}                                               & 53.1*                                             & 450*                                                                       & 18*                                                                   & 1.4*                                                                    \\ \hline
\end{tabular}
\label{Table:2}
\end{table}

\subsection{Comparison and Discussion}
Table \ref{Table:2} shows a comparison of our SiN-on-SiO$_2$ MRR modulator with the simulation (marked as *) and fabrication based best-performing nine SiN MRR modulators from prior works (\cite{phare2015graphene}-\cite{hermans2019integrated},\cite{wang2022}-\cite{wang2022ultra},\cite{zhao2022}), in terms of five key attributes namely optical modulation bandwidth (O-MB), electrical modulation bandwidth (E-MB), modulation efficiency (ME), FSR, and energy-efficiency (EE). The SiN MRR modulator in \cite{phare2015graphene} achieves higher O-MB compared to the other SiN MRR modulators (Table \ref{Table:2}) and our modulator. In contrast, our modulator achieves higher E-MB compared to the other SiN MRR modulators (Table \ref{Table:2}). We have also evaluated that our modulator achieves the best effective MB of $\sim$46.2 GHz compared to all other SiN MRR modulators, based on the formula of effective MB from \cite{dubray2016}. Due to its superior effective MB of $\sim$46.2 GHz, our modulator can be easily operated at $>$15 Gb/s bitrate to enable ultra-high-speed (potentially beyond Tb/s) DWDM-based PICs while ensuring minimal power-penalty from crosstalk \cite{bahadori2016crosstalk}. Moreover, our modulator achieves higher ME compared to other SiN MRR modulators (Table \ref{Table:2}). However, in terms of FSR, SiN MRR modulator demonstrated in \cite{liu2020} achieves higher FSR compared to the other SiN MRR modulators in Table \ref{Table:2} including our modulator. Nevertheless, our modulator consumes the energy of 1.4 pJ/bit which is significantly better than the energy consumption of the modulator from \cite{liu2020}. Its high energy efficiency and competent FSR of 18 nm make our modulator a favorable candidate for designing high-bandwidth and energy-efficient DWDM-based photonic interconnects for datacenter-scale as well as chip-scale computing and communication architectures. 


Further, although ITO is not available in the CMOS process flow, it can be deposited at relatively low temperatures (less than 300°C) on top of the back-end-of-line (BEOL) metal layers of CMOS chips, in an independent manner without interfering with or contaminating the CMOS front-end-of-line (FEOL) and BEOL processes. This makes our SiN-on-SiO$_2$ modulator an excellent choice for implementing optical interconnect PICs on silicon interposers, to enable ultra-high-bandwidth inter-chiplet communication in emerging multi-chiplet systems \cite{Popstar}.

\section{Conclusion}
We have demonstrated an ITO-based SiN-on-SiO$_2$ MRR modulator, which consists of a stack of ITO-SiO$_2$-ITO thin films as the active upper cladding of the SiN MRR core. This active upper cladding of our modulator leverages the free-carrier assisted, high-amplitude refractive index change in the ITO films to affect a large electro-refractive optical modulation in the device. To evaluate the performance of our SiN-on-SiO$_2$ MRR modulator, we performed electrostatic, transient, and FDTD simulations using the foundry-validated Ansys/Lumerical tools. Based on these simulations, our modulator achieves superior performance with 450 pm/V modulation efficiency, $\sim$46.2 GHz 3-dB modulation bandwidth, 18nm FSR, 0.24 dB insertion loss, and 8.2 dB extinction ratio for OOK modulation at 30 Gb/s. This excellent performance of our SiN-on-SiO$_2$ MRR modulator demonstrates its potential to enhance the performance and energy efficiency of SiN-on-SiO$_2$ based PICs of the future. 

\begin{backmatter}
\bmsection{Disclosures} The authors declare no conflicts of interest
\end{backmatter}

\bibliography{ref}

\bibliographyfullrefs{ref}

\end{document}